\documentclass[prb,preprint]{revtex4-2}
\usepackage[dvipdfmx]{graphicx}
\usepackage{dcolumn}
\usepackage{bm}
\usepackage{color}
\usepackage{here}
\usepackage{braket}
\usepackage{amsmath,amssymb}
\newcommand{\ham}{\hat{\mathcal{H}}}
\newcommand{\hamd}{\hat{\mathcal{H}}^\prime}

\newcommand{\kp}{{\bm k \cdot \bm p}}
\newcommand{\Ek}{\mathcal E - \bm k}
\newcommand{\SIAk}{\mathrm{SIA}{\bm k}}
\newcommand{\cSOI}{\mathrm{cSOI}}
\newcommand{\SIAp}{\mathrm{SIA}{\bm p}}

\newcommand{\BIAk}{\mathrm{BIA}{\bm k}}
\newcommand{\BIAp}{\mathrm{BIA}{\bm p}}

\newcommand{\sige}{\mathrm{Si}_{0.5}\mathrm{Ge}_{0.5}}
\begin{document}

\title{Non-adiabatic Berry phase for semiconductor heavy holes under the coexistence of Rashba and Dresselhaus spin--orbit interactions}

\author{Tatsuki Tojo}
\email{tojo@qms.cache.waseda.ac.jp}
\author{Kyozaburo Takeda}
\email{takeda@waseda.jp}
\affiliation{
Faculty of Science and Engineering, 
Waseda University,
Shinjuku, Tokyo, 169-8555, Japan
}
\date{\today}

\begin{abstract}
We formulate the non-Abelian Berry connection (tensor $\mathbb R$) and phase (matrix $\bm \Gamma$) for a multiband system and apply them to semiconductor holes under the coexistence of Rashba and Dresselhaus spin--orbit interactions.
For this purpose, we focus on the heavy-mass holes confined in a $\sige$ two-dimensional quantum well, whose electronic structure and spin texture are explored by the extended $\kp$ approach.
The strong intersubband interaction in the valence band causes {\it quasi}-degenerate points except for point $\Gamma$ of the Brillouin zone center. 
These points work as the singularity and change the Abelian Berry phase by the quantization of $\pi$ under the adiabatic process.
To explore the influence by the non-adiabatic process, we perform the contour integral of $\mathbb R$ faithfully along the equi-energy surface by combining the time-dependent Schr\"{o}dinger equation with the semi-classical equation-of-motion for cyclotron motion and then calculate the energy dependence of $\bm \Gamma$ computationally.
In addition to the function as a Dirac-like singularity, the {\it quasi}-degenerate point functions in enhancing the intersubband transition via the non-adiabatic process. 
Consequently, the off-diagonal components generate both in $\mathbb R$ and $\bm \Gamma$, and the simple $\pi$-quantization found in the Abelian Berry phase is violated.
More interestingly, these off-diagonal terms cause ``resonant repulsion'' at the {\it quasi}-degenerate energy and result in the discontinuity in the energy profile of $\bm \Gamma$.
\end{abstract}

\maketitle

\section{Introduction}

Semiconductors have a valence band, comprising three types of subbands: heavy-mass holes (HHs), light-mass ones (LHs), and split-off ones (SHs). 
An existence of these subbands causes intersubband interaction (ISI), which produces anisotropy and non-parabolicity in the valence band, and then the subbands have {\it quasi}-degenerate states in the Brillouin zone (BZ).
Around the {\it quasi}-degenerate point, one can obtain the eigenvectors that are mutually orthogonal but cannot determine them uniquely due to a ``phase" indefiniteness. 
Moreover around the point, the spin--orbit interaction (SOI) induced by Dresselhaus (bulk-inversion-asymmetry, BIA)~\cite{Dres} and/or Rashba (structure-inversion-asymmetry, SIA)~\cite{Rashba} terms is expected to generate the characteristic spin texture via the complicated distribution of the effective magnetic field~\cite{ganichev,PLA,pssb}.
Consequently, Berry's analysis~\cite{berry} is required for semiconductor holes when these {\it quasi}-degenerate states have the $\Ek$ structure, similar to the Dirac singularity~\cite{massive}. 

Considering the above background, we studied the energy and temperature dependence of the Berry phase for HHs confined in a $\sige$ two-dimensional quantum well (2DQW) where the SIA and BIA SOIs coexist~\cite{PLA,pssb}. 
To investigate the ISI, we extended the Dresselhaus, Kip, and Kittel (hereafter, abbreviated DKK)~\cite{DKK} $\kp$ perturbation approach by including second-order crossings between the $\kp$ and SOI terms and determined the periodic part of the $n$-th Bloch hole ($\ket{u^n_{\bm k}}$).
Our extended $\kp$ approach demonstrates that the {\it quasi}-degenerate points resemble the 2D Dirac singularity~\cite{massive}. Those points in the $\Braket{110}$ direction work as the ``monopole"-like singularity, whereas those in the $\Braket{\bar{1}10}$ direction work as a ``dipole"-like singularity. Consequently, the Berry phase of HH has the unique energy-dependent plateau with $\pi$ quantization. Moreover, this $\pi$ quantization is understandable by counting the number of {\it quasi}-degenerate points in the BZ alongside recognizing the sign of the Berry curvature in the area~\cite{PLA,pssb}. 

We elucidated that the {\it quasi}-degenerate state in the valence band surely has the potential to work as the Dirac singularity. However, our treatment seems to be one-sided because the calculation was performed under the adiabatic process. We ignored the intersubband transition although the {\it quasi}-degenerate point provides a small energy difference between the different-type holes.
One should consider the influence of the non-adiabatic process via the intersubband hybridization.
Therefore, we here formulate the non-Abelian Berry connection (tensor $\mathbb R$) and phase (matrix $\bm \Gamma$) for the multiband system (Sec. \ref{theotre}).
Several pioneering studies have focused on the non-Abelian treatment of Berry's analysis, and they have deepened our understanding, particularly in the field of the quantum information and computation~\cite{Wilczek,pachos, wysokinski, weisbrich, Zu2014}.
Here, we developed our formulation in conformity with the practical subject of semiconductor holes, sacrificing mathematical strictness, 
because we wish to apply our formulation to semiconductor holes under the coexistence of Rashba and Dresselhaus SOIs in Sec. \ref{SiGeapp}.
We focus on HHs confined in $\sige$ 2DQW, whose electronic structure and spin texture are explored by the extended $\kp$ approach (Subsec. \ref{extkp}).
We then calculate the energy dependence of $\bm \Gamma$ computationally by performing the contour integral of $\mathbb R$ faithfully along the equi-energy surface by combining the time-dependent (TD) Schr\"{o}dinger equation with the semi-classical equation-of-motion for the cyclotron motion (Subsec. \ref{Berrysec}).
We studied the energy dependence of the Berry phase by dividing the energy into the characteristic regions such as the lower state and the state around the {\it quasi}-degenerate point and then explore the influence by the non-adiabatic process.

\section{Theoretical treatment for non-adiabatic processes}
\label{theotre}

\subsection{Non-Abelian Berry connection tensor}
\label{rateandnonabelian}

The Berry phase is the phase difference of the wave function and is defined by the contour integral when the final state exactly turns back to the initial state during the specific motion characterized by the parameters describing the system accurately.
Because we explore the Berry phase of the carrier having the Bloch state, we can use a wave vector $\bm k$ as the parameter describing the system well. We define the periodic part of the $n$-th Bloch eigenstate by $\ket{u_{\bm k}^n}$.
Thus, any state $\ket{\phi_{\bm k}}$ of the carrier is represented by the expansion of $\ket{u_{\bm k}^n}$ because of the completeness at $\bm k$:
\begin{align}
\label{tenkai}
\ket{\phi_{\bm k}}=\displaystyle \sum_n c^n_{\bm k}\ket{u^n_{\bm k}},
\end{align}
where $n$ includes the spin polarization.

Now, we explore the $\bm k$-space trajectory whose final state $\bm k_f$ completely returns to the initial state $\bm k_i$. 
The equi-energy surface is the representative of this trajectory, and the ``cyclotron motion" realizes it. Thus, we remake the ``dynamical" process of the carrier by cyclotron motion, whose TD feature is described by the semi-classical equation of motion.
This remaking by the cyclotron motion enables us to solve the time dependence of the wave vector by coupling the semi-classical equation of motion with the TD Schr\"{o}dinger equation.
Accordingly, we can explicitly include the time dependence in the projection coefficient $c^n_{\bm k}$ in Eq. \eqref{tenkai}, although the wave vector $\bm k$ itself is inherently time-independent. 
Additionally, energy conservation during the cyclotron motion leads to the following physical insights:
The adiabatic process specifically determines the equi-energy surface of the single eigenstate $\ket{u^n_{\bm k}}$.
Consequently, the closed trajectory with $\bm k_i = \bm k_f$ is naturally and uniquely determined by the single-cycle motion. 
Contrary, the non-adiabatic process allows the carrier to cause interstate hybridization, even during cyclotron motion along the energy-conserved $\bm k$-space trajectory.
The resulting $\bm k$-space trajectory cannot be represented by the single equi-energy surface of the specific eigenstate; however, it is represented by synthesizing the multi-equi-energy surfaces because the carrier goes back and forth among the multiple states.
Thus, the single-cycle cyclotron motion in the non-adiabatic process does not always equalize the final wave vector $\bm k_f$ to the initial one $\bm k_i$.
Multicycle cyclotron motion might be required to create the closed trajectory of $\bm k_i = \bm k_f$.

By employing Eq. \eqref{tenkai}, we rewrite the TD Schr\"{o}dinger equation as follows:
\begin{align}
\label{tdeq}
i\hbar\frac{d}{dt} \ket{\phi_{\bm k}(t)} = \ham_{\bm{k}} \ket{\phi_{\bm k}(t)},
\end{align}
into the following rate equation:
\begin{align}
\label{rate}
\frac{d c^m_{\bm k}}{dt}&= - \displaystyle{ \sum_{m^\prime}
\left ( \sum_\xi \Braket{ u^m_{\bm k} | \frac{\partial u^{m^\prime}_{\bm k}}{\partial k_\xi}}
\frac{dk_{\xi}}{dt}
\right )} c^{m^\prime}_{\bm k} - i \frac{E^m_{\bm k}}{\hbar}c^{m}_{\bm k} \nonumber \\
&=i \displaystyle{ \sum_{m^\prime}
\left ( \sum_\xi \mathbb R^\xi_{m m^\prime}(\bm k)\
\dot{k}_{\xi}
\right )} c^{m^\prime}_{\bm k} - i \frac{E^m_{\bm k}}{\hbar}c^{m}_{\bm k}.
\end{align}
Here, we define a tensor component $\mathbb R^\xi_{m m^\prime}(\bm k)$ as follows: 
\begin{align}
\label{Rtensor}
\mathbb R^\xi_{m m^\prime}(\bm k)
=i \Braket{ u^m_{\bm k} | \frac{\partial u^{m^\prime}_{\bm k}}{\partial k_\xi}}.
\end{align}
Because we focus on the 2D system in this work, we focus on the in-plane components $\xi=x$ and $y$. This procedure does not remove the generality of the formulation \eqref{Rtensor}.

The tensor $\mathbb R^\xi_{m m^\prime}(\bm k)$ defined by Eq. \eqref{Rtensor} is mathematically equivalent to the Berry connection tensor.
Physically, this tensor is the non-Abelian Berry connection for the non-adiabatic process~\cite{Wilczek,CandN} because it considers the interstate hybridization between $\ket{u^m_{\bm k}}$ and $\ket{u^{m^\prime}_{\bm k}}$.
The non-commutativity of the non-Abelian Berry connection tensor appears explicitly in the time-ordered product in the calculation of the Berry phase matrix.

The diagonal term ($m^\prime=m$) of Eq. \eqref{Rtensor} is given as follows: 
\begin{align}
\label{Am1}
\mathbb R_{mm}^{\xi}(\bm{k})=i\Braket{u_{\bm{k}}^m|\frac{\partial u_{\bm{k}}^{m}}{\partial k_{\xi}}} \equiv {A}^{\xi}_{m}(\bm{k}).
\end{align}
The symbol ${A}^{\xi}_{m}(\bm{k})$ is the $\xi$ component of the Abelian Berry connection vector $\bm{A}_m(\bm{k})$ for the $m$-th carrier, which is given by
\begin{align}
\label{Am2}
\bm{A}_m(\bm{k})=i\bra{u_{\bm{k}}^m}\nabla_{\bm{k}}\ket{u_{\bm{k}}^m}.
\end{align}

\subsection{Non-Abelian Berry phase matrix}

To eliminate the dynamical phase from $\mathbb R^\xi_{m m^\prime}(\bm k)$, we further redefine the rationalized Berry connection tensor $\bar{\mathbb R}^\xi (\bm k)$ as follows:
\begin{align}
\label{Rp}
\bar{\mathbb R}^\xi (\bm k)
=\left( \exp \left[ i\int _0 ^t \bm \Xi _{\bm k (t^\prime)} dt^\prime / \hbar \right] \right )
\mathbb R^\xi (\bm k)
\left( \exp \left[ -i\int _0 ^t \bm \Xi _{\bm k (t^\prime)} dt^\prime / \hbar \right ] \right ).
\end{align}
Here, we introduce the energy eigenvalue matrix $\bm \Xi _{\bm k}$, whose diagonal elements are those eigenvalues of the Bloch eigenstates:
\begin{align}
\label{Ekdiag}
\bm \Xi _{\bm k}=
\begin{pmatrix}
E^1_{\bm k} &&\text{\raisebox{-10pt}[0pt][0pt]{\hspace{-10pt}\huge{0}}} \\
& E^2_{\bm k} \\
\text{\raisebox{5pt}[0pt][0pt]{\hspace{5pt}\huge{0}}}& & \ddots
\end{pmatrix}.
\end{align}
Accordingly, we have the non-Abelian Berry phase matrix $\bm \Gamma (E)$ when the final point $\bm k_f(E)$ ($t=T_{\mathrm{cls}}$) completely returns to the initial point $\bm k_i(E)$ ($t=0$) by
\begin{align}
\label{BG}
\bm \Gamma (E) 
= - i \ln \left( \mathcal{T} \exp \left[ i \displaystyle \sum_\xi \int _0 ^ {T_c} \bar{\mathbb R}^\xi \ \dot{k_\xi}(E) dt \right] \right )
= - i \ln \left( \mathcal{T} \exp \left[ i \displaystyle \sum_\xi \int _{\bm k_i(E)} ^ {\bm k_f(E)} \bar{\mathbb R}^\xi \ {dk_\xi(E)} \right] \right ).
\end{align}
Here, we define the time derivation of the wave vector $\bm k$ by the symbol $\dot{\bm k}(=d \bm k/dt)$.
We also employ the time-order product $\mathcal{T}$ considering the non-commutativity in the Berry connection tensor $\displaystyle{\bar{\mathbb R}^\xi (\bm k(E))}$.

If the system is well described by the adiabatic approximation, the off-diagonal elements (non-commutative) of the rationalized Berry connection $\bar{\mathbb R}$ are negligible, leading to the diagonal Berry connection $\bm R_m (=\bm A_m)$. Accordingly, $\mathbb R$ at any time is commutative, and the time-ordered product simply offsets the mathematical operations `$\ln$' and `$\exp$' in Eq. \eqref{BG}. 
Consequently, Stokes' theorem transforms the contour integral to the surface integral as
\begin{align}
\label{apprx1}
\bm \Gamma = 
- i \ln \left( \exp \left[ i \displaystyle \int _{\bm k_i(E)} ^ {\bm k_f(E)} \bar{\mathbb R} \cdot {d \bm k(E)} \right] \right ) 
\Rightarrow \oint _E \bm R \cdot d\bm k = \iint_{\leq E} \nabla \times \bm R \ dk_x dk_y.
\end{align}

\subsection{Cyclotron motion}

To calculate the non-Abelian Berry phase, we must perform the contour integral of Eq. \eqref{BG} by paying particular attention to the time-ordered product caused by the non-commutativity of the non-Abelian Berry connection. 
Beside this attention, we are required to describe the time dependence of the wave vector because the integrand has the term $\dot{\bm k}$.
The magnetic field $\bm B_0=(0, 0, B_0)$ applied to resolve the spin degeneracy leads to ``cyclotron motion", by which we can remake the motion of the wave vector along the $\bm k$ trajectory.
The semi-classical equation of motion describes this TD feature as follows:
\begin{align}
\label{semicla}
\frac{d \bm{k}}{dt}=\frac{q}{\hbar^2}\displaystyle{
\Bra{\phi_{\bm k}}\nabla_{\bm{k}}\ham_{\bm k}\Ket{\phi_{\bm k}}\times {\bm B}_0
}.
\end{align}
Accordingly, the coupling of equations \eqref{rate} and \eqref{semicla} enables us to carry out the contour integral along the cyclotron $\bm k$-trajectory accurately by considering the time-ordered product in Eq. \eqref{BG}.

\section{Application to $\sige$ binary alloy system}
\label{SiGeapp}

\subsection{extended $\kp$ approach and {\it quasi}-degenerate states}
\label{extkp}

We have extended the $\kp$ approach by considering the crossings between the $\kp$ and SOI couplings up to the second (2nd) order terms to study the spin textures of the Rashba and Dresselhaus SOI competition system~\cite{PLA,pssb}. 
The extended $\kp$ approach determines both the periodic part $\ket{u^n_{\bm k}}$ and the eigen energy $E_{\bm k}^n$ of the $n$-th Bloch hole $\ket{\varphi_{\bm k}^n ({\bm r})}=e^{i\bm k \cdot \bm r} \ket{{u_{\bm k}^n(\bm r)}}$, precisely taking into account the ISI:
\begin{eqnarray}
\label{blocheq}
\ham_{\bm{k}}^{\mathrm{exd}}\ket{u_{\bm k}^n}=\left( \ham_0 + \hamd_{\bm{k}} + \mu_B\hat{ \bm \sigma} \cdot \bm B_0 \right)\ket{u_{\bm k}^n}
=\left( E_{\bm k}^n- \frac{\hbar^2}{2m}k^2 \right)\ket{u_{\bm k}^n}
\equiv \mathcal{E}_{\bm k}^n\ket{u_{\bm k}^n}.
\end{eqnarray}
Here, $\ham_0$ is the non-perturbed Hamiltonian, and $\hamd_{\bm{k}}$ is the extended $\kp$ Hamiltonian given by 
\begin{eqnarray}
\label{Hd}
\hamd_{\bm{k}}
= \hamd_{\kp} + \hamd_{\cSOI}
+ \hamd_{\SIAk}+\hamd_{\SIAp \otimes \kp}
+ \hamd_{\BIAk}+\hamd_{\BIAp \otimes \kp}.
\end{eqnarray}
Here, $\hamd_{\kp}$ is the conventional 2nd-order $\kp$ perturbation Hamiltonian introduced by DKK~\cite{DKK}, and $\hamd_{\cSOI}$ is the internal SOI perturbed one caused by the crystal potential (corresponding to so-called $\bm \ell \cdot \bm s$). The remaining four terms in Eq. \eqref{Hd} correspond to the SOI terms related to the SIA and BIA, as defined in our previous work~\cite{jpsj0,jpcm}. 

In the calculation of the electronic structure, holes are supposed to be confined in the Si$_{0.5}$Ge$_{0.5}$ binary alloy system having an alternating configuration where the BIA is most strengthened. 
We further suppose the 2DQW system having 41 atomic layers, whose thickness is realistic and reproduces the above BIA influence.
We apply the Rashba electric field ($\Xi_0 = 1.04 \times 10^{3}$ [m/s]) perpendicular to the (001) quantum plane to study the SIA-BIA coexistence system~\cite{pssb,wolos,zhang2020,studer,reine}. 
Furthermore, we apply a very weak magnetic field (4.7 mT) perpendicular to the quantum plane. The smallness of this magnetic field hardly changes the electronic structure and spin texture from those under the zero magnetic field but resolves the degeneracy at not only point $\Gamma$ (BZ center) but also the {\it quasi}-degenerate points ($\braket{110}$ and $\braket{\bar{1}10}$), imitating them to the massive Dirac Fermion.

\begin{figure}[h]
\begin{center}
\includegraphics[scale=0.5]{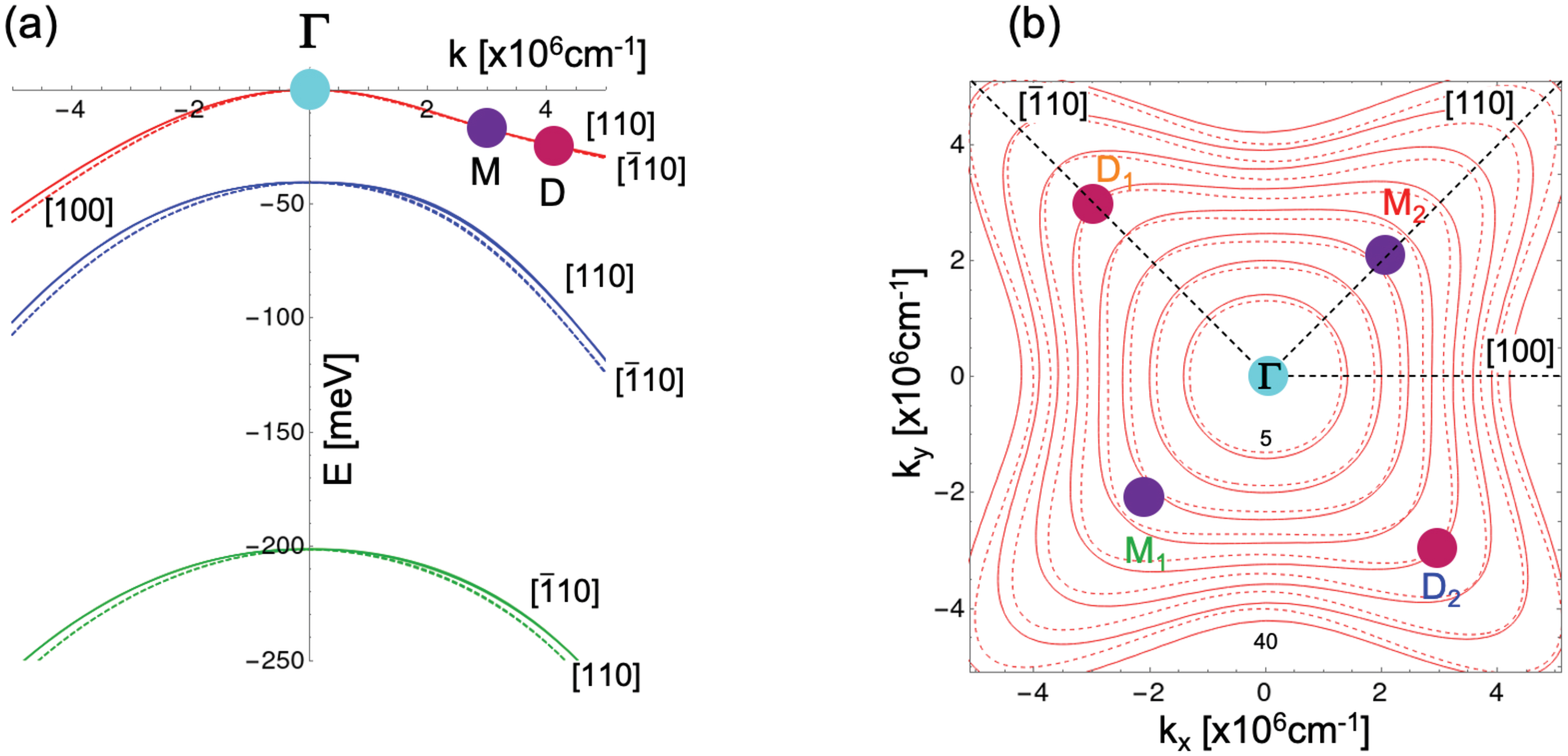}
\caption{The $\Ek$ dispersion relation (a) and equi-energy contours (b) of the 2DQW system consisting of Si$_{0.5}$Ge$_{0.5}$ alternating alloy. The system has the (001) quantum plane and has the equivalent strength between SIA and BIA couplings. The $\Ek$ dispersion relations of HHs, LHs, and SHs are red, blue, and green, respectively; the solid lines represent states with a stabilized spin (+), whereas the broken ones are those having a destabilized spin ($-$). We show the equi-energy contours of HH$\pm$ with an energy interval of 5 meV (b). The colored circles in the figures indicate the {\it quasi}-degenerate points between HH$\pm$. All $\Ek$ dispersions (a) and equi-energy surfaces (b) are those of the ground state against the 2D quantization.
}
\label{fig1}
\end{center}
\end{figure}

Figure \ref{fig1}(a) shows the resultant $\Ek$ dispersion relations of HHs, LHs, and SHs confined in the $\sige$ 2DQW system under the SIA-BIA coexistence~\cite{PLA}. 
Because the SOIs caused by the Rashba and Dresselhaus couplings stabilize($+$)/destabilize($-$) each type of holes (HHs, LHs, and SHs), we identify them by solid/broken lines. 
We then show the corresponding equi-energy contours of HH$\pm$ in Fig. \ref{fig1}(b).
The ISI causes the strong anisotropy and non-parabolicity both in HH$\pm$ toward the $\braket{110}$ directions, leading to {\it quasi}-degeneracy between them, as found in Fig. \ref{fig1}(b). The $C_2$ symmetry of the system results in the two pairs of the {\it quasi}-degenerate points: $M_1$ and $M_2$ in the $[\bar{1}\bar{1}0]$ and [110] directions, respectively, and $D_1$ and $D_2$ in the $[1\bar{1}0]$ and $[\bar{1}10]$ directions, respectively.
As reported in our previous works~\cite{PLA,pssb}, the former {\it quasi}-degenerate points M work as a ``monopole" singularity, whereas the latter ones D work as a ``dipole" one.

\subsection{non-Abelian Berry phase}
\label{Berrysec}

\begin{figure}[htbp]
\begin{center}
\includegraphics[scale=0.4]{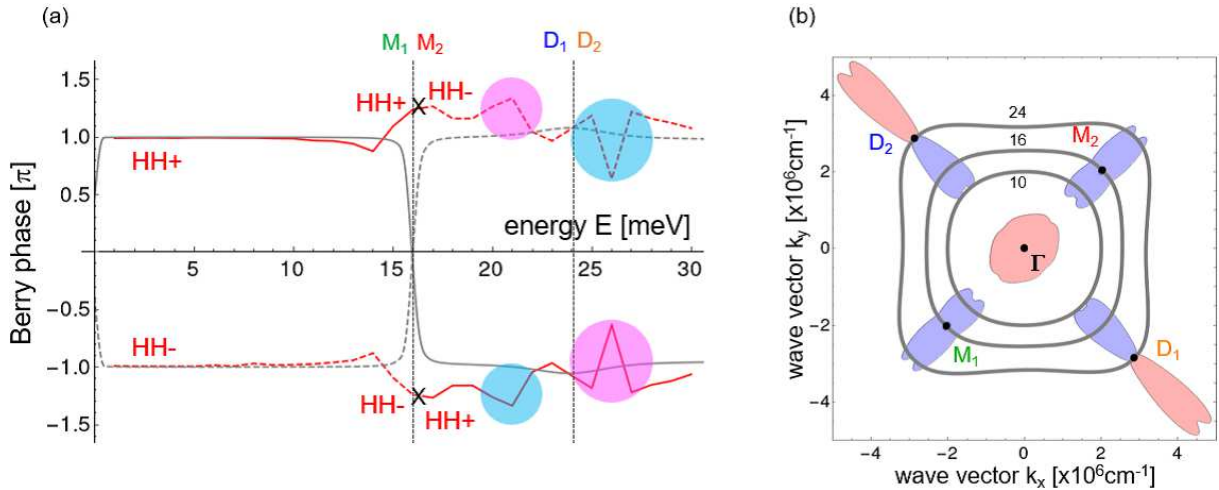}
\caption{Energy dependence of the non-Abelian Berry phase for HH$+$ and HH$-$ beyond the adiabatic approximation (a). For a comparison, we overwrite the corresponding energy-dependence of the Abelian Berry phase (gray lines). Stokes' theorem enables us to calculate the Abelian Berry phase by the surface integral of the Abelian Berry curvature for HH+, whose distribution is illustrated in Figure (b). 
}
\label{fig2}
\end{center}
\end{figure}

By employing the non-Abelian Berry connection $\mathbb R$, we explore the energy dependence of the Berry phase $\bm \Gamma$ for HH$\pm$ beyond the adiabatic process. 
After the diagonalization of Eq. \eqref{blocheq}, we store the eigenstates $\ket{u^n_{\bm k}}$ and $E^n_{\bm k}$ beforehand. Then, we call and employ them for the numerical calculation at each $\bm k(t)$, in accordance with the time proceeding. 
The coupling of the semiclassical equation of motion for the cyclotron motion \eqref{semicla} with the rate equation \eqref{rate} enables us to rewrite the TD cyclotron motion into the $\bm k$ space trajectory and introduce the time dependence implicitly in the eigen vectors along the trajectory.
In the calculation, we set the initial state of the considering hole $\ket{\phi_{\bm k}}$ to the eigen state of HH+ having $\bm k_i^{\mathrm {HH+}}=(k_i^{\mathrm {HH+}}, 0)$, and perform the contour integral along the $\bm k$-space cyclotron trajectory under the energy conservation.
The time-ordered product in the integration of Eq. \eqref{BG} is carried out by the Crank-Nicolson algorithm.
In the practical calculation, we divide the $\bm k$ space into the line element of $dk \sim 1\times 10^{3} \mathrm{cm}^{-1}$. 
Accordingly, the time-ordered product along the equi-energy surface (e.g., $E=10$ meV) requires the calculation of 13,550,000 steps. 
We executed the present numerical calculations with an accuracy within an error of $1\times 10^{-9}\ \%$ of the total energy to confirm the energy conservation during the ``cyclotron motion''. 
Due to the matrix form of the non-Abelian Berry phase $\bm \Gamma$ (Eq. \eqref{BG}), we diagonalize it and identify those eigenvalues of HH$\pm$, LH$\pm$ and SH$\pm$, based on the components, $\ket{yz}$, $\ket{zx}$ and $\ket{xy}$ including a spin. 

Figure \ref{fig2}(a) shows the eigenvalues ${\gamma}_{\mathrm{HH}\pm}(E)$ against energy. The solid (red) line indicates the energy dependence of HH+, whereas the broken (red) line does that of HH$-$.
Similar to other semiconductors, the present $\sige$ 2DQW system has a massive Dirac singularity at point $\Gamma$, and the Berry phase of $\pm \pi$ results in HH$\pm$, irrespective of the adiabatic/non-adiabatic processes.
Figure \ref{fig2}(a) demonstrates that the non-Abelian Berry phase ${\gamma}(E)$ for HH$\pm$ also converges into $\pm \pi$ toward the valence band top.
Because the Abelian Berry phase in the adiabatic process is determined by the singularity of the massive/massless Dirac point, HH$\pm$ maintains the value $\pm \pi$ until the {\it quasi}-degenerate points M$_1$ and M$_2$ newly appear.
In the lower energy region with an energy of less than $\sim 12$ meV, the non-adiabatic process hardly changes the energy dependence in $\gamma(E)$ from that found in the adiabatic process; i.e., the non-adiabatic process gives a similar plateau profile with $\gamma_{\text{HH}\pm}(E)= \pm\pi$.
However, at around 16 meV, the non-adiabatic process unexpectedly changes the sign of the Berry phase and causes a discontinuity in $\gamma(E)$ by interchanging mutually. This feature is completely different from the abrupt but continuous change in the sign of the Berry phase found in the adiabatic process.
One should further note that the non-adiabatic process causes the characteristic ``bumpy" profile around 21 and 26 meV, being different from the flat profile ($\gamma=\mp \pi$) in the adiabatic process.

\begin{figure}[htbp]
\begin{center}
\includegraphics[scale=0.35]{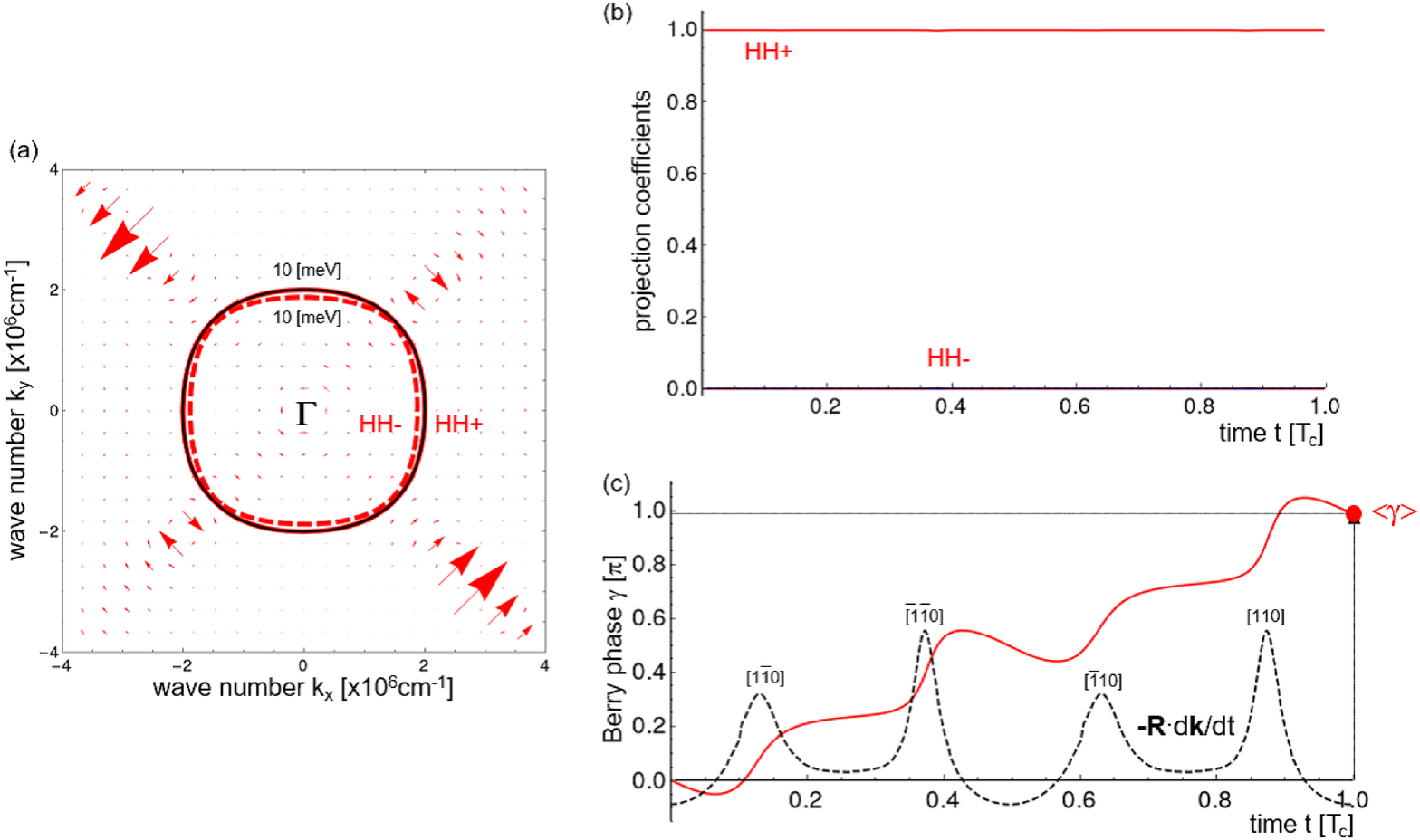}
\caption{$\bm k$-space trajectory (black solid line) of the hole having an energy of 10 meV (a). We illustrate the vector-field distribution of the non-Abelian Berry connection (red arrows) for HH$+$ with the equi-energy surfaces (10 meV) of HH+ (red solid line) and HH$-$ (red broken line). 
We show the projection profile (b), and the successively integrated values (red line) of the inner product (black broken line) between $\bar{\mathbb R}$ and $\dot{\bm k}$ (c).
}
\label{fig3}
\end{center}
\end{figure}

\subsubsection*{in the lower energy region}

We start exploring the non-Abelian Berry phase ${\gamma}(E)$ up to $\sim 10$ meV.
Figure \ref{fig3}(a) shows the vector field of the in-plane Berry connection.
We also illustrate the $\bm k$ trajectory for the cyclotron motion while overwriting the equi-energy surfaces of HH$\pm$ with 10 meV.
The cyclotron motion trajectory completely coincides with the equi-energy surface of HH+. Moreover, the final position $\bm k_f$ of the hole accurately returns to the initial position $\bm k_i$ by the single cycle motion.
The projection analysis reveals that the hole $\ket{\phi_{\bm k}}$ consists of HH+ mostly, and less states are hybridized even under the non-adiabatic process during the cyclotron motion.
Accordingly, the $\bm k$-space trajectory coincides with the HH+ equi-energy surface of 10 meV (Fig. \ref{fig3}(b)).

Figure \ref{fig3}(c) shows the successive integration (red line) of the inner product between the Berry connection ${\mathbb R}$ and the time-derivative of the wave vector $\dot{\bm k}$. We perform integration until  time $t$ stepwise along the cyclotron trajectory. Any values at midway are {\it gauge variant} and meaningless physically, except for the full contour integral. However, the ``midway" integrals well explain the determination process of the Berry phase.
Figure \ref{fig3}(a) demonstrates that the non-Abelian Berry connection causes the large values in the $\braket{110}$ and $\braket{1\bar{1}0}$ directions. 
Moreover, those connection vectors are anti-parallel to the cyclotron direction of the hole having 10 meV.
Consequently, the cosine value between $\bar{\mathbb R}$ and $\dot{\bm k}$ increases stepwise every time the hole passes those four directions (broken line in Fig. \ref{fig3}(c)). 
Except for the above four directions, the value of the Berry connection vector itself is small and the minute inner product is obtained.
One should also note that the system has the $C_2$ rotational symmetry, by which the hole repeats the same TD profile on the ``integrated" one when the time passes half of the cycle time $T_c$ ($=T_{\mathrm{cls}}/N_{\mathrm{cyc}}$, where $T_{\mathrm{cls}}$ and $N_{\mathrm{cyc}}=1$ are the time and cycles required to form the closed trajectory ($\bm k_i = \bm k_f$), respectively). 
In such a way, the hole in the non-adiabatic process passes through these complicated {\it gauge-variant} processes. 
Nevertheless, the non-Abelian Berry phase ${\gamma}$ determined by the contour integral interestingly coincides with the value under the adiabatic process.
That is, the Berry phase of HH+ (and HH$-$) up to $\sim$12 meV is well described by the adiabatic approximation. 

\subsubsection*{on the monopole point M}

\begin{figure}[htbp]
\begin{center}
\includegraphics[scale=0.3]{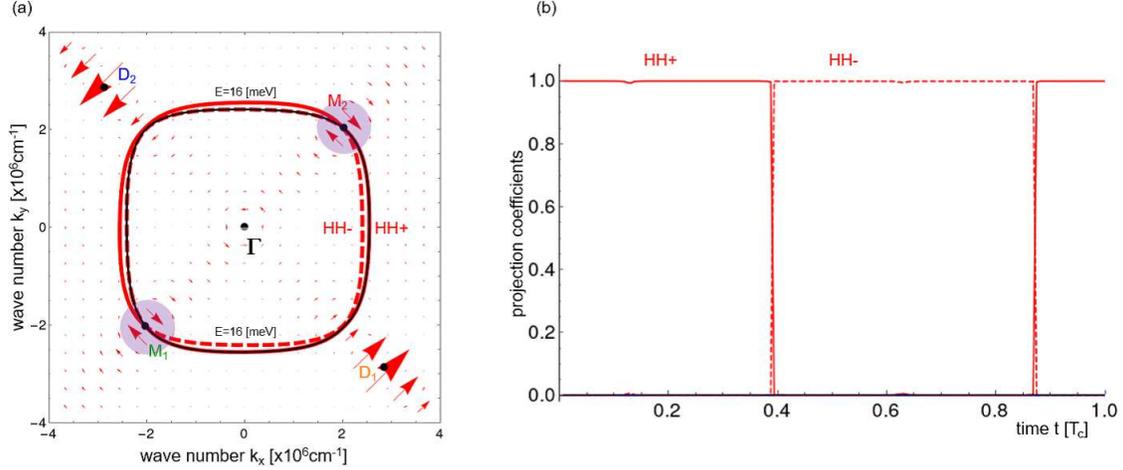}
\caption{$\bm k$-space trajectory (black solid line) of the hole having an energy of 16 meV with the vector-field distribution of the non-Abelian Berry connection (red arrows) and the equi-energy surfaces (16 meV) of HH+ (red solid line) and HH$-$ (red broken line) (a). 
We show the projection coefficients $|c_{\bm k}^m|^2$ against time (b).}
\label{fig4}
\end{center}
\end{figure}

We explore how the non-Abelian Berry phase is affected when the hole passes the {\it quasi}-degenerate points M$_1$ and M$_2$. For this purpose, we set the hole initially with the HH+ eigenstate having 16 meV.
The resulting $\bm k$ trajectory and the projection coefficients are shown in Fig. \ref{fig4}(a) and (b), respectively.
The hole first illustrates the $\bm k$ trajectory equal to the equi-energy surface of HH+ (red solid line) because the hole initially has the HH+ eigenstate. 
Figure \ref{fig4}(b) demonstrates that the hole interchanges completely from HH+ to HH$-$ when it passes point M$_1$. Accordingly, the hole changes the trajectory from the equi-energy surface of HH$+$ (red solid line) to that of HH$-$ (red broken line).
The $C_2$ symmetry of the system further causes the transition back to HH+ at the {\it quasi}-degenerate point M$_2$. 
Consequently, the trajectory returns to the equi-energy surface of HH+, and the closed trajectory having $\bm k_i = \bm k_f$ is produced inevitably by the single cyclotron cycle.
Thus, the hole interchanges completely and alternately between HH$\pm$ states every time when it passes the ``monopole" points M$_1$ and M$_2$, and goes back and forth over the two equi-energy surfaces of HH$\pm$ during the cyclotron motion due to energy conservation.

\begin{figure}[htbp]
\begin{center}
\includegraphics[scale=0.4]{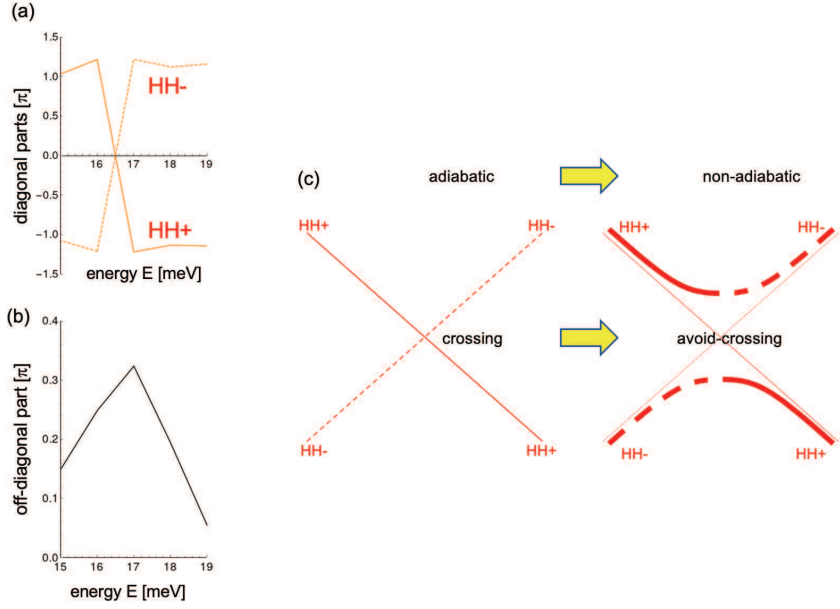}
\caption{Energy dependence of the non-Abelian Berry phase matrix for HH$\pm$ around the {\it quasi}-degenerate points M; diagonal elements (a) and off-diagonal elements (b). 
Schematic explanation for the discontinuity in the energy dependence of the non-Abelian Berry phase around the {\it quasi}-degenerate points M (c).
}
\label{fig5}
\end{center}
\end{figure}

The non-adiabatic process allows the interstate transition, and then induces the off-diagonal terms $\mathbb R^\xi_{mm^\prime}(\bm k)$.
Consequently, the non-Abelian Berry connection vector has the characteristic distribution with the vortex and the opposite orientation at the {\it quasi}-degenerate point M, as shown in Fig. \ref{fig4}(a). 
This off-diagonal Berry connection element $\mathbb R^\xi_{mm^\prime}(\bm k)$ further induces the off-diagonal terms of the non-Abelian Berry phase matrix $\bm \Gamma$.
Figure \ref{fig5} illustrates the energy dependence of the diagonal (a) and off-diagonal (b) elements of $\bm \Gamma$ for HH$\pm$ around the {\it quasi}-degenerate point M.
The absolute value of the diagonal element decreases and becomes zero at point M. 
Then, those diagonal terms cross mutually and increase. The energy profiles of these diagonal elements resemble those Abelian Berry phases, where the adiabatic process prohibits the interstate hybridization (Fig. \ref{fig5}(a)).
In contrast, the non-adiabatic process leads to the off-diagonal terms having the maximum at point M (Fig. \ref{fig5}(b)).
That is, the intercrossing of the Berry phases at point M changes into ``avoid crossing" by the off-diagonal elements which are caused by the interstate transition via the non-adiabatic process. 
This ``resonant repulsion" causes discontinuity in the non-Abelian Berry phases for HH$\pm$ at the {\it quasi}-degenerate point M (Fig. \ref{fig2}(a)). 
Moreover, the interstate transition breaks the rigorous $\pi$-quantization. 

\begin{figure}[htbp]
\begin{center}
\includegraphics[scale=0.35]{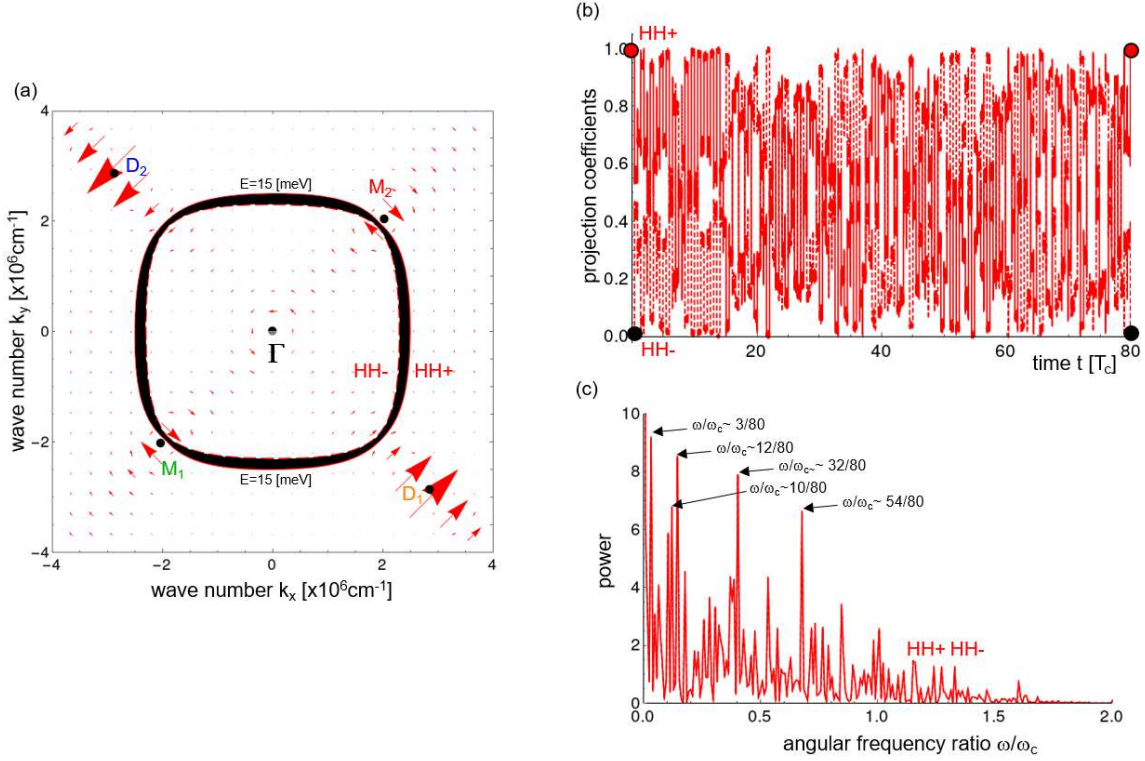}
\caption{$\bm k$-space trajectory (black solid line) of the hole with an energy of 15 meV, far from the {\it quasi}-degenerate energy by 1 meV (a). We overwrite the vector-field distribution of the non-Abelian Berry connection (red arrows) and the equi-energy surfaces (15 meV) of HH+ (red solid line) and HH$-$ (red broken line). 
We show the projection coefficients $|c_{\bm k}^m|^2$ against time (b) and their Fourier components (c).}
\label{fig6}
\end{center}
\end{figure}

What happens when the cyclotron trajectory is apart just from point M?
Figure \ref{fig6} shows the $\bm k$ trajectory (a), the projection coefficients (b), and their Fourier analysis (c) for the hole having 15 meV.
More apart from point M, more a larger energy difference between HH$\pm$ ($\Delta E^\pm$) results. 
Consequently, the interstate transition between HH$\pm$ is reduced (Fig. \ref{fig6}(b) and (c)). 
Nevertheless, the $\bm k$ trajectory (a) is formed by synthesizing the equi-energy surfaces of HH$+$ and HH$-$ having 15 meV in accordance with the hybridization ratio.
The non-adiabatic process results in this complicated interstate transition and requires 80 cyclotron cycles to close the trajectory ($\bm k_f(E) =\bm k_i(E)$).

We suppose that the formation of the closed trajectory ($\bm k_i = \bm k_f$) requires $N_{\mathrm{cyc}}$ cycles as well as time $T_{\mathrm{cls}}$. 
Then, we get the {\it pseudo}-cyclotron period and -frequency by $T_c=T_{\mathrm{cls}}/N_{\mathrm{cyc}}$ and $\omega_c=N_{\mathrm{cyc}} \cdot \omega_{\mathrm{cls}}$, where the symbol $\omega_{\mathrm{cls}}=2\pi/T_{\mathrm{cls}}$ is the fundamental frequency.
Therefore, the $l$-th harmonics has the frequency of $\omega_l= l \cdot \omega_{\mathrm{cls}}$, and the frequency ratio against the pseudo-cyclotron frequency is a rational number, given by,
$$ \frac{\omega_l}{\omega_{c}}= \frac{l}{N_{\mathrm{cyc}}}.$$
The Fourier analysis (Fig. \ref{fig6}(c)) demonstrates that any peaks are expressed by a rational number with a denominator of 80.

\subsubsection*{on the dipole point D}

\begin{figure}[htbp]
\begin{center}
\includegraphics[scale=0.3]{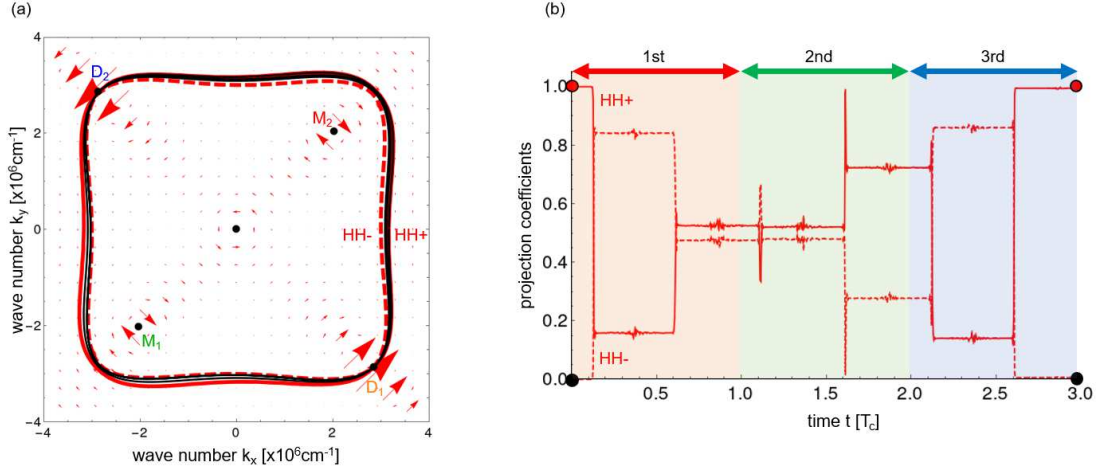}
\caption{$\bm k$-space trajectory (black solid line) of the hole having the {\it quasi}-degenerate energy of 24 meV with overwriting the non-Abelian Berry connection vector field (red arrows) and the equi-energy surfaces (24 meV) of HH+ (red solid line) and HH$-$ (red broken line) (a). 
We show the projection coefficients $|c_{\bm k}^m|^2$ against time (b). We color the first, second, and third cycles with red, green, and blue, respectively.}
\label{fig7}
\end{center}
\end{figure}

We further explore the $\bm k$ trajectory (a) and the projection profile (b) when the hole passes other {\it quasi}-degenerate points D$_1$ and D$_2$ around 24 meV (Fig. \ref{fig7}).
Despite the small energy difference between the states HH$\pm$ at point D$_1$, the transition into HH$-$ state is not full, different from the situation at point M$_1$. 
The projection profile further demonstrates that the hole consists of both states of HH$\pm$ during the cyclotron motion.
Accordingly, the resulting trajectory is between the equi-energy surfaces of HH+ and HH$-$; e.g., from the point D$_1$ to D$_2$ in the second cycle, the hole has nearly an even ratio of the hybridization between HH$\pm$, and the trajectory is at the midway between two equi-energy surfaces of HH$\pm$.
When the hole passes point D$_2$, the non-full transition to HH+ enforces the hole not to return its trajectory to the equi-energy surface of HH+. 
Consequently, the cyclotron trajectory cannot be closed by the single cycle and multiple cycles with three cycles are required.

\begin{figure}[htbp]
\begin{center}
\includegraphics[scale=0.3]{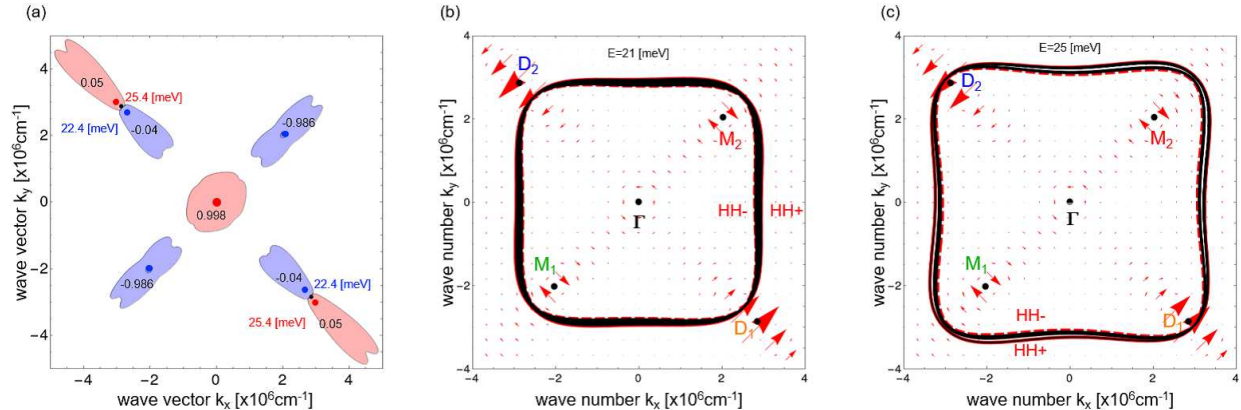}
\caption{$\bm k$-space distribution of the Abelian Berry curvature for HH+ (a), and the resulting cyclotron trajectory (black solid line) of the hole having the {\it quasi}-degenerate energy of 21 meV (b) and 25 meV (c). We illustrate the contour of the Abelian Berry curvature of $\pm$0.001 in (a) and overwrite the vector field of the non-Abelian Berry connection (red arrows) and the equi-energy surfaces of HH+ (red solid line) and HH$-$ (red broken line) in (b) and (c). 
}
\label{fig8}
\end{center}
\end{figure}

We reinvestigate the distribution of the Abelian Berry curvature near the ``dipole"-like singularity D (Fig. \ref{fig8}(a)). 
In the $[1\bar{1}0]$ direction, we find the minimum point of the negative curvature at 22.4 meV, whereas the maximum point of the positive one is at 25.4 meV.
We carry out surface integration of the curvature surrounded by the contour illustrated in Fig. \ref{fig8}(a); the negative area has a value of $-0.04\pi$ and the positive area $0.05\pi$.
Thus, those values of $-0.04\pi$ and $0.05\pi$ cancel the surface integral over point D, resulting in the ``dipole"-like nature. 
Consequently, this weakened singularity results in the ``convex and dent" of the non-Abelian Berry phases, found around 21 and 25 meV in Fig. \ref{fig2}(a).
In contrast, the corresponding surface integration around point $\Gamma$ and point M has values of $0.998\pi$ and $-0.986\pi$, respectively, and these points function as the Dirac singularity.

We illustrate the $\bm k$ trajectories having an energy of 21 and 25 meV in Fig. \ref{fig8}(b) and (c).
One should remember that the cyclotron motion passing on the ``monopole"-like point M requires a single or a few cycles to close the cyclotron trajectory due to full-alternate transition.
However, the above negative (21 meV) and positive (25 meV) minimum points deviate from point D, and the energy difference between HH$\pm$ increases.
Accordingly, the TD components of the hole interchange neither fully nor alternately due to reducing interstate transitions between these two states.
Thus, closing of the trajectory requires the multi-times cyclotron motion, such as 19 (b) and 7 times (c), respectively.

\section{Summary}

We formulated the non-Abelian Berry connection tensor $\mathbb R$ and phase matrix $\bm \Gamma$ for the multiband system and applied them to semiconductor holes under the coexistence of Rashba and Dresselhaus SOIs.
We, then, calculated the energy dependence of $\bm \Gamma$ computationally, focusing on HHs confined in $\sige$ 2DQW. 
We carried out the contour integral of $\mathbb R$ along the equi-energy surface by combining the TD Schr\"{o}dinger equation with the semi-classical equation-of-motion for the cyclotron motion.

The ISI causes the {\it quasi}-degenerate states plurally in the semiconductor valence band. These {\it quasi}-degenerate points work as a Dirac's singularity and cause $\pi$ quantization in the energy dependence of the Berry phase under the adiabatic process. 
The non-adiabatic process induces intersubband hybridization, and the off-diagonal elements both in $\mathbb R$ and $\bm \Gamma$ increase particularly around the {\it quasi}-degenerate points. 
Consequently, the simple $\pi$-quantization in the Berry phase is violated.
More interestingly, the non-Abelian Berry phase for HH$\pm$ interchanges mutually at the {\it quasi}-degenerate energy of point M. As such, HH$\pm$ has the discontinuity in the energy dependence of $\bm \Gamma$. 
This interchange-and-discontinuity is explainable by the non-adiabatic process, through which the off-diagonal terms are generated and the ``resonant repulsion" is formed.

\appendix

\section{Non-Abelian Berry curvature}
\label{secomega12}

By employing the 2D non-Abelian Berry connection tensor $\mathbb R^{\xi}_{m m^\prime}$ of Eq. \eqref{Rtensor}, the out-of-plane component of the non-Abelian Berry curvature matrix $\Omega^z_{m m^\prime}$ is given by
\begin{align}
\label{Omegatensor}
\Omega^z_{m m^\prime}({\bm k})
&=\left [ \nabla_{\bm k} \times \mathbb R_{m m^\prime}({\bm k}) \right]_z
-i \displaystyle \sum_{l} \left[\mathbb R_{m l}({\bm k}) \times \mathbb R_{l m^\prime}({\bm k}) \right]_z \nonumber \\
&=\left ( \displaystyle{ \frac{\partial \mathbb R_{m m^\prime}^y({\bm k})}{\partial k_x} - \frac{\partial \mathbb R_{m m^\prime}^x({\bm k})}{\partial k_y}} \right )
-i \sum_{l} \left ( \mathbb R_{m l}^x({\bm k}) \mathbb R_{l m^\prime}^y({\bm k}) 
- \mathbb R_{m l}^y({\bm k}) \mathbb R_{l m^\prime}^x({\bm k}) \right )\nonumber \\
&\equiv \Omega_{mm^\prime}^{(1)z}(\bm{k}) +\Omega_{mm^\prime}^{(2)z}(\bm{k}).
\end{align}
Here, we define the symbols $\Omega_{mm^\prime}^{(1)z}(\bm{k})$ and $\Omega_{mm^\prime}^{(2)z}(\bm{k})$ by
\begin{align}
\label{omega12}
\Omega_{mm^\prime}^{(1)z}(\bm{k})&\equiv \left[\nabla_{\bm k} \times \mathbb R_{m m^\prime}({\bm k}) \right]_z 
=\frac{\partial \mathbb R_{mm^\prime}^{y}(\bm{k})}{\partial k_x}-\frac{\partial 
\mathbb R_{mm^\prime}^{x}(\bm{k})}{\partial k_y}, \nonumber \\
\Omega_{mm'}^{(2)z}(\bm{k}) &\equiv-i\sum_{l}\left[\mathbb R_{ml}(\bm{k})\times \mathbb R_{lm'}(\bm{k})\right]_z \nonumber \\
&=-i\sum_{l}\left[{\mathbb R}_{ml}^x(\bm{k}){\mathbb R}_{lm'}^y(\bm{k})-{\mathbb R}_{ml}^y(\bm{k}){\mathbb R}_{lm'}^x(\bm{k})\right].
\end{align}

\begin{figure}[htbp]
\begin{center}
\includegraphics[scale=0.45]{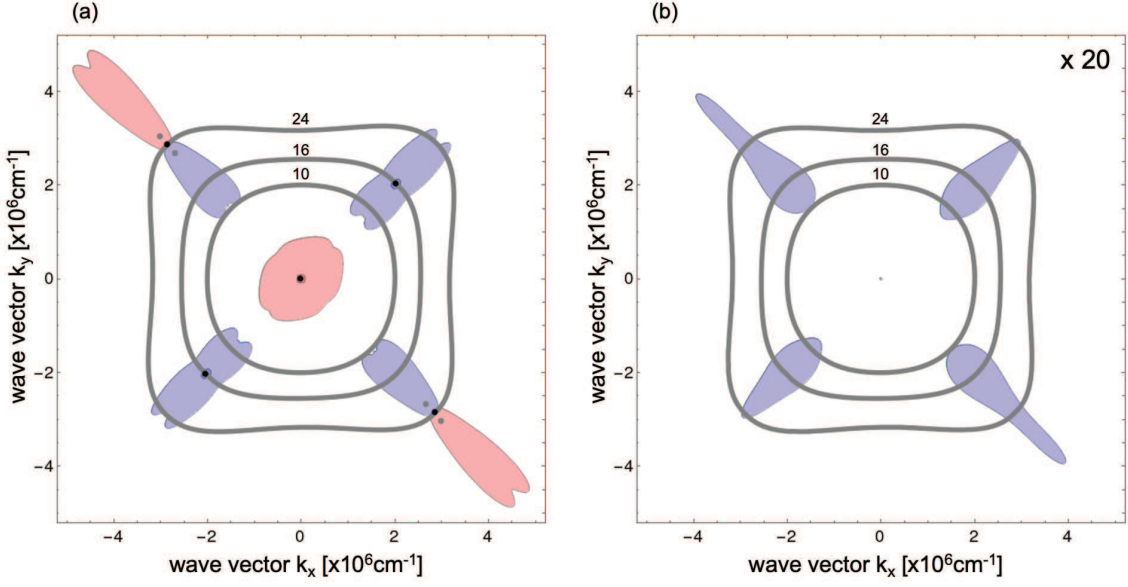}
\caption{$\bm k$ space distribution of the in-plane diagonal component of the non-Abelian Berry curvature for HH+; $\Omega_{mm}^{(1)z}(\bm{k})$ (a) and $\Omega_{mm}^{(2)z}(\bm{k})$ (b).}
\label{fig9}
\end{center}
\end{figure}

Because the non-Abelian Berry curvature has a matrix form $\Omega_{mm'}^{z}(\bm{k})$ due to the non-adiabatic process, we cannot compare the non-Abelian result with the Abelian one straightforwardly. We therefore focus on the diagonal elements of $\Omega_{mm}^{(1)z}(\bm{k})$ and $\Omega_{mm}^{(2)z}(\bm{k})$ and explore the influence by the non-adiabatic process via the diagonal elements.
Figures \ref{fig9}(a) and (b) show the diagonal elements of $\Omega_{mm}^{(1)z}(\bm{k})$ and $\Omega_{mm}^{(2)z}(\bm{k})$, respectively.
The comparison of Fig. \ref{fig9}(a) with Fig. \ref{fig8}(a) demonstrates that the diagonal component $\Omega_{mm}^{(1)z}(\bm{k})$ coincides with the Abelian Berry curvature. 
The diagonal element $\Omega_{mm}^{(1)z}(\bm{k})$ is rewritten into 
\begin{align}
\Omega_{mm}^{(1)z}(\bm{k})&=\frac{\partial \mathbb R_{mm}^{y}(\bm{k})}{\partial k_x}-\frac{\partial \mathbb R_{mm}^{x}(\bm{k})}{\partial k_y}
=\frac{\partial A_m^{y}(\bm{k})}{\partial k_x}-\frac{\partial A_m^{x}(\bm{k})}{\partial k_y}
=B_m^z(\bm{k}).
\end{align}
Here, symbol $A_m^{\xi}(\bm{k})$ is the Abelian Berry connection for $m$-th hole, as defined in Eq. \eqref{Am1}.
As such, the diagonal element $\Omega_{mm}^{(1)z}(\bm{k})$ exactly gives the out-of-plane ($z$) components of the Abelian Berry curvature, leading to complete coincidence between Figs. \ref{fig9}(a) and \ref{fig8}(a).

We also have the diagonal element $\Omega_{mm}^{(2)z}(\bm{k})$ of
\begin{align}
\Omega_{mm}^{(2)z}(\bm{k})&=-i\sum_{l}\left[{\mathbb R}_{ml}^x(\bm{k}){\mathbb R}_{lm}^y(\bm{k})-{\mathbb R}_{ml}^y(\bm{k}){\mathbb R}_{lm}^x(\bm{k})\right]
=2\sum_{l}\mathrm{Im}\left[{\mathbb R}_{ml}^x(\bm{k}){\mathbb R}_{lm}^y(\bm{k})\right].
\end{align}
Here, $l$ sums over the states {\it outside} of the HH space~\cite{CandN}.
Figure \ref{fig9}(b) demonstrates that the resulting diagonal element $\Omega_{mm}^{(2)z}(\bm{k})$ is 1/20th that of $\Omega_{mm}^{(1)z}(\bm{k})$.
Particularly in the lower energy region at energies less than 10 meV, we get $\Omega_{mm}^{(2)z}(\bm{k}) \sim 0$. This is the reason why the non-Abelian Berry phase in such a lower energy is approximated by the adiabatic process. 
However, the non-adiabatic process causes intersubband hybridization with those {\it outside} states, especially around the {\it quasi}-degenerate points M and D, and results in 
the small but finite values in the diagonal component of the Berry curvature.
\begin{align}
\label{totalomega}
\Omega_{mm}^{z}(\bm{k})=\Omega_{mm}^{(1)z}(\bm{k})+\Omega_{mm}^{(2)z}(\bm{k})
=B_m^z(\bm{k})+\Omega_{mm}^{(2)z}(\bm{k}).
\end{align}
That is, the non-adiabatic process gives the influence {\it even on} the diagonal component of the non-Abelian Berry curvature via $\Omega_{mm}^{(2)z}(\bm{k})$.

\end{document}